# Quantum teleportation in high dimensions

Yi-Han Luo, Han-Sen Zhong, Manuel Erhard, Xi-Lin Wang, Li-Chao Peng, Mario Krenn,
Xiao Jiang, Li Li, Nai-Le Liu, Chao-Yang Lu[*], Anton Zeilinger, and Jian-Wei Pan[#]

*Hefei National Laboratory for Physical Sciences at Microscale and Department of Modern Physics,*
*University of Science and Technology of China, Hefei, 230026, China,*
*CAS Centre for Excellence in Quantum Information and Quantum Physics, Hefei, 230026, China,*
*Austrian Academy of Sciences, Institute for Quantum Optics and Quantum Information (IQOQI),*
*Boltzmanngasse 3, A-1090 Vienna, Austria,*
*Faculty of Physics, University of Vienna, Boltzmanngasse 5, 1090 Vienna, Austria*
\* cylu@ustc.edu.cn, # pan@ustc.edu.cn

**Abstract:**

**Quantum teleportation allows a "disembodied" transmission of unknown quantum states between distant quantum systems. Yet, all teleportation experiments to date were limited to a two-dimensional subspace of quantized multiple levels of the quantum systems. Here, we propose a scheme for teleportation of arbitrarily high-dimensional photonic quantum states, and demonstrate an example of teleporting a qutrit. Measurements over a complete set of 12 qutrit states in mutually unbiased bases yield a teleportation fidelity of 0.75(1), which is well above both optimal single-copy qutrit-state-estimation limit of 1/2 and maximal qubit-qutrit overlap of 2/3, thus confirming a genuine and non-classical three-dimensional teleportation. Our work will enable advanced quantum technologies in high dimensions, since teleportation plays a central role in quantum repeaters and quantum networks.**

The laws of quantum mechanics forbid precise measurement or perfect cloning of unknown quantum states (*1*). With the help of shared entanglement and classical communication channel, however, quantum teleportation in principle allows faithful transfer of the unknown quantum states from one particle to another at a distance (*2*), without physical transmission of the object itself. There have been numerous experiments on the teleportation of quantum states of single photons (*3-8*), atoms (*9*), trapped ions (*10, 11*), defects in solid states (*12*), and superconducting circuits (*13*). All these quantum systems naturally possess not only multiple degrees of freedom, but also, many degrees of freedom can have high quantum numbers (*14*) beyond the simplified two-level qubit subspace. However, all experiments to date were limited to two-dimensional subspaces, the qubits (*3-13*). After the teleportation of two-particle composite state (*6*) and two degrees of freedom (*7*), the teleportation of high-dimensional quantum states remained the final obstacle for teleporting a quantum particle intact.

The ability of coherent control of high-dimensional (HD) quantum states is important for developing advanced quantum technologies. Compared to the conventional two-level systems, HD states can offer extended possibilities such as both higher capacity and noise resilience in quantum communications (*15, 16*), more efficient quantum simulation (*17*) and computation (*18*), as well as larger violation of Bell inequality (*19*). Recent years have witnessed an increasing capability to generate and measure HD entangled states (*19-24*). However, the previous work is predominantly limited in the coherent control of single particle HD state. A joint projection of two independent particles with unknown states into maximally entangled HD states, which requires some forms of controlled interactions and will play a crucial role in the HD teleportation, dense coding, and quantum computing, is more challenging and remains largely unexplored experimentally.

Here, we propose an efficient and extendable scheme for teleportation of arbitrarily HD photonic quantum states, and we report the first experimental teleportation of a qutrit, which is equivalent to a spin-1 system. We start by describing our protocol of HD quantum teleportation. For the sake of simplicity, here we explain it using the example of a three-level system, where the underlying physics can be generalized to arbitrary *N*-level systems (*25*), ~~as elaborated in~~

~~detail in the supplementary materials~~. Suppose Alice wishes to teleport to Bob the quantum state

$$| \varphi \rangle_a = \alpha_0 | 0 \rangle_a + \alpha_1 | 1 \rangle_a + \alpha_2 | 2 \rangle_a \qquad [1]$$

of a single photon, where $| 0 \rangle$, $| 1 \rangle$, and $| 2 \rangle$ are encoded by three different paths of the photon (c.f. Fig. 1), their subscripts label the photon, and their coefficients are complex numbers that fulfil $|\alpha_0|^2 + |\alpha_1|^2 + |\alpha_2|^2 = 1$. The quantum resource required for teleporting this state is a HD entangled state previously shared between Alice and Bob, for example,

$$| \psi_{00} \rangle_{bc} = (| 00 \rangle_{bc} + | 11 \rangle_{bc} + | 22 \rangle_{bc}) / \sqrt{3} \,. \qquad [2]$$

This is one of the three-dimensional (3D) Bell states which, together with other eight orthogonal ones (*25*), forms a complete orthonormal basis of the bipartite 3D Hilbert space.

Conceptually within the theoretical framework of Bennett *et al.* (*2*), the most crucial step for the HD teleportation is performing a joint measurement of photon *a* and *b*, a process called 3D Bell-state measurement (BSM). With equal probabilities of 1/9, the 3D-BSM projects photon *a* and *b* into one of the nine 3D Bell states randomly. Alice can then broadcast the 3D-BSM result classically, which allows Bob to accordingly apply a unitary single-particle 3D transform (*26*) to reconstruct the original quantum states (eqn.1) at his location (*25*). In general, for an *N*-level bipartite system, there exists $N^2$ HD Bell states. An unambiguous HD-BSM poses a new challenge both theoretically and experimentally.

Recall that in the teleportation of qubits, the four Bell states can be grouped into three symmetric and one anti-symmetric state under particle exchange, which facilitate the discrimination using linear optics (*27*). In the 3D case already, however, the situation becomes fundamentally more complicated. There are three Bell states that are symmetric and the other six are neither symmetric nor anti-symmetric. In theory, it was shown (*28*) that it is impossible to discriminate two-photon HD Bell states with linear optics only when the dimensions $N \geq 3$.

To overcome such a linear optical limitation (*29*), here we utilize *N*-2 additional single photons, so-called ancillary photons, and a multiport beam splitter (BS) with *N*-input-*N*-output all-to-all connected ports (*29*), which is a generalisation of the quantum Fourier transform. A detailed derivation of the HD-BSM procedure is shown in (*25*).

To get a deeper insight on how the high-dimensional Bell-state measurement works, we use the fact that in reverse a Bell state is generated. This simplifies the analysis because in this case we can focus on one specific "click-pattern" as an example, and send single photons backwards from these detectors. We choose to propagate three indistinguishable single photons from the output ports $\{a_0^{'}, a_1^{'}, a_2^{'}\}$ backwards through the multi-port, as shown in Fig. 1. Then we condition onto cases where in each input port one and only one photon exists. Due to the all-to-all connection in the multi-port, the resulting state contains all length-2 permutations of the 3D states (25). Detection of the ancillary photon in the superposition state $(|0\rangle_x + |1\rangle_x + |2\rangle_x)/\sqrt{3}$, results in the obtained state (normalization omitted):

$$|0\rangle_a(|1\rangle_b + |2\rangle_b) + |1\rangle_a(|0\rangle_b + |2\rangle_b) + |2\rangle_a(|0\rangle_b + |1\rangle_b) \qquad [3]$$

The unitary transformation of this state to a target 3D Bell state (Eqn.2) requires an expanded Hilbert space of four dimensions. The extra fourth level $|3\rangle$, is added to assist the physical realization of the unitary transformation (see $U_{3+1}$ in Fig. 1), and erased afterwards, which leads to the target 3D Bell state (Eqn. 2) $|\psi_{00}\rangle_{ab} = (|00\rangle_{ab} + |11\rangle_{ab} + |22\rangle_{ab})/\sqrt{3}$ (25). The analysis holds exactly the same if the three indistinguishable photons are "incident" from ports $\{b_0^{'}, b_1^{'}, b_2^{'}\}$ or $\{x_0^{'}, x_1^{'}, x_2^{'}\}$.

Thus, in the experiment, a simultaneous click of the three detectors in the ports $\{a_0^{'}, a_1^{'}, a_2^{'}\}$, $\{b_0^{'}, b_1^{'}, b_2^{'}\}$ or $\{x_0^{'}, x_1^{'}, x_2^{'}\}$ indicates an unambiguous projection of the input photons *a* and *b* to the 3D Bell state $|\psi_{00}\rangle_{ab}$, which projects Bob's photon *c* onto $|\varphi\rangle_c = \alpha_0 |0\rangle_c + \alpha_1 |1\rangle_c + \alpha_2 |2\rangle_c$.

This state is already identical to the original state of photon $a$ without the need of any additional unitary corrections. The success probability of the HD-BSM using this scheme is 1/81. Combining with active feed-forward techniques increases the success probability to 1/9 for linear optics.

Figure 2 shows the experimental set-up for the 3D quantum teleportation. A femtosecond pulsed laser beam is split into two parts to simultaneously create two photon pairs. The first part of the pump beam is divided into three paths by two beam-displacers which are then focused on the same β-borate-borate (BBO$_1$) crystal. We select the case where in total one photon pair is produced by type-II beam-like spontaneous parametric down-conversion (*30*) but without knowing at which one of the three paths, which generates the desired entangled state $|\psi_{00}\rangle_{bc}$ used as the quantum channel for the 3D teleportation. To ensure long-term phase stability between the three paths, we specifically design and fabricate interferometers with small (4 mm) separation between the three paths. Hence, air fluctuations and disturbances act collectively on all paths such that the qutrits are effectively protected in a decoherence-free space.

Before being sent to the HD-BSM, photon $b$ from the entangled qutrits first undergoes the unitary transformation $U_{3+1}$, which is experimentally realized using a network of polarising beam splitters and half-wave plates (HWPs). The details are shown in supplementary materials (*25*). Another pump beam from the same laser passes through BBO$_2$ and creates the second photon pair. One of them is used for the preparation of an arbitrary superposition of the three paths as the input state $a$ to be teleported. The other one is used as the ancillary qutrit $x$ in the HD-BSM.

The experimental realization of the HD-BSM puts stringent technological requirements on phase stability, efficiency, and precision. To meet these demands, the HD-BSM is operated in hybrid polarisation-path encoding and employs a fully connected three-input and three-output ultra-low-loss multi-port interferometer. As shown in Fig. 2, in the input, single photons $a$ and $b$ are initialised in horizontal polarization, while photon $x$ is in vertical polarization. Photon $a$ is first combined with photon $x$ using a PBS. The combined beams pass through a HWP set at $22.5^0$, and are superposed with photon $b$ on a partially-polarisation-dependent beam splitter (pPDBS) which totally reflects vertically polarised photons and partially—with a ratio of 1/3—reflects horizontally polarised photons. One of the output ports of the pPDBS is detected by three detectors directly, while the other port is further sent through a quarter-wave plate set at $45^0$ and then split by a polarising beam splitter and detected by six detectors. It is straightforward to check that all the three photons from the inputs $a$, $b$, and $x$ are evenly distributed to each of the output with a ratio of 1/3, realizing the most important function of the multiport.

The all-to-all multiport in Fig. 2 involves three-photon nine-path Hong-Ou-Mandel interferences at the polarising beam splitter and the pPDBS. All nine paths are synchronized to arrive within ~10 fs of each other, a delay much smaller than the coherence time of the narrowband (3 nm) filtered single photons (~450 fs). The use of compact and precisely aligned beam-displacers ensures a good spatial overlap in all multi-path interferences simultaneously. Verifications of all two-photon Hong-Ou-Mandel interference combinations at the pPDBS with an average visibility of 0.82(1) are presented in supplementary materials (25). These visibilities in combination with the entanglement source (with a measured fidelity of 0.94(1)) quantify the quality of the three-dimensional teleportation experiment (25). ~~see supplementary materials for details.~~

It is necessary to prove that the teleportation experiment works universally for all possible superposition states in the general form of $|\varphi\rangle_a$ (eqn. [1]) and has a performance exceeding that using only classical methods. Classically, the optimal single-copy state-estimation fidelity of a three-level quantum system (*31*) is 0.5 when averaging over the whole Hilbert space. Sampling only over partial state space in biased bases, however, would allow the classical strategy to make use of the biased information to obtain an average state estimation fidelity higher than 0.5. It is therefore important to carefully choose a minimal set of input states such that the random sampling of which leads to the same classical limit as sampling over the whole state space. Such a minimal set of states lies in mutually unbiased bases (*32*). For a three-dimensional system, we need to measure 12 states from four mutually unbiased bases:

$$
\begin{aligned}
B^{(1)}_{\{1,2,3\}} &: (1,0,0),\ (0,1,0),\ (0,0,1) \\
B^{(2)}_{\{1,2,3\}} &: (1,1,1),\ (1,\omega,\omega^2),\ (1,\omega^2,\omega) \\
B^{(3)}_{\{1,2,3\}} &: (\omega,1,1),\ (1,\omega,1),\ (1,1,\omega) \\
B^{(4)}_{\{1,2,3\}} &: (\omega^2,1,1),\ (1,\omega^2,1),\ (1,1,\omega^2)
\end{aligned}
$$

[4]

Here the vectors $(\alpha_0, \alpha_1, \alpha_2)$ denote the state $\alpha_0 |0\rangle_c + \alpha_1 |1\rangle_c + \alpha_2 |2\rangle_c$, $\omega = \exp(i2\pi/3)$, where the normalising constant is omitted. We measure fidelities of the final teleported states, defined as the overlap of the experimentally measured density matrix $\rho_c$ with the ideal input state $|\psi\rangle_a$, which can be written as $\mathrm{Tr}(|\psi\rangle\langle\psi|\rho)$. Conditioned on the detection of a specific "click-pattern" within the HD-BSM (see Fig. 1), we register the counts of Bob's photon and analyse its properties. The verifications of the teleportation results are based on four-fold coincidence detections which in our experiment occur with a rate of ~0.11 Hz. In each setting, the typical data accumulation time is 20-40 minutes, which allows to sufficiently suppress Poisson noise.

Figure 3**A** shows the teleportation results of group $B^{(1)}_{\{1,2,3\}}$, which can be straightforwardly measured in the computational basis. The extracted fidelities are 0.76(3), 0.81(3), and 0.78(3) for the teleported state $|0\rangle$, $|1\rangle$, and $|2\rangle$, respectively. However, the measurements for the other three groups, which involves equal superpositions of all the three levels, are more complicated. To experimentally access the fidelity of these states in the general form

$$|\phi\rangle = (|0\rangle + \exp(i\phi_1)|1\rangle + \exp(i\phi_2)|2\rangle)/\sqrt{3} \ , \qquad [5]$$

we decompose the density matrix into three parts, $\rho = (\sigma_{012} + \sigma_{021} + \sigma_{120})/3$, where

$$\sigma_{ijk} = |\varphi_{ij}^+\rangle\langle\varphi_{ij}^+| - |\varphi_{ij}^-\rangle\langle\varphi_{ij}^-| + |k\rangle\langle k| \ ,$$

$$|\varphi_{ij}^{\pm}\rangle = (|i\rangle \pm \exp(i\phi_j - \phi_i)|j\rangle)/\sqrt{2} \ .$$

The decomposition unitarily transforms the qutrits into two-dimensional superposition states and one computational state. Our measurement apparatus allows a simultaneous three-outcome readout, directly accessing one of the $\sigma_{ijk}$ (*25*). We show in Fig. 3**B**, **C**, and **D** the measurement results for three representative states from the group of $B_{\{1,2,3\}}^{(2)}$, $B_{\{1,2,3\}}^{(3)}$, and $B_{\{1,2,3\}}^{(4)}$, respectively. The other six states are presented in Figure S6. We note that all reported measurements are without background or accidental count subtraction. The fidelity imperfection is mainly from double pair emissions, spatial mode mismatch in the multi-photon multi-path interferences, and interferometric noise in the state preparation and measurements (*25*).

The fidelities of all the 12 states are displayed in Fig. 4, which are the minimal set allowing us to faithfully derive the teleportation fidelity for the three-level quantum system. In the current experiment, the averaged fidelity is calculated to be 0.75(1), well above the classical limit of 0.5 which can be obtained with the best classical strategy.

Proving the universality and non-classicality is already sufficient for teleporting qubits. However, for the *N*-dimensional teleportation, it is important to further verify that all *N*-dimensions still can form a coherent superposition and thus survived the teleportation intact. Hence, a genuine *N*-dimensional teleportation should be distinguished from lower-dimensional cases by excluding the hypotheses that the teleported state could be represented with less dimensions. For our specific 3D teleportation, we can calculate that the maximal overlap between any two-level superposition and the genuine three-level states is 2/3 (*25*). The teleportation fidelity measured in our work exceeds this threshold by 9 standard deviations, thus conclusively establishing a genuine three-dimensional quantum teleportation.

To summarize, we have for the first time demonstrated the possibility to completely teleport the multiple quantized levels of a quantum system. Our generalized scheme (*25*) can readily be applied to other degrees of freedom, such as photon's orbital angular momentum (*33*). Future implementations with higher dimensions would be suitable to be implemented in integrated photonics platforms (*22,24*). It would also be interesting to investigate in the future the teleportation of multiple atomic levels in trapped ions (*10,11*) and cold atoms (*9*). An intriguing idea appears upon combining our approach with the teleportation of a two-particle composite state (*6*) and two degrees of freedom (*7*), which makes it possible to realise the dream of teleporting a complex quantum system completely.

The ability to perform the HD-BSM developed in this work provides a new possibility for fundamental test of Bell's inequalities and advanced quantum information technologies. As an example, the to-be-teleported HD quantum state can itself be fully undefined, such as being part of a two-particle HD entanglement. This leads to entanglement swapping (*34*), where heralded by a HD-BSM click, two remote HD particles can be entangled with no direct interaction. Such a scheme can distribute entanglement over long distances and can enable an event-ready Bell test. Remarkably, the created HD entanglement can tolerate a higher detection inefficiency than the qubit case (*19, 35*), and would provide significant advantages in long-distance Bell test closing both locality and detection loopholes (*36-38*) and device-independent quantum key distribution (*39*).

Note added: After completing our work, we became aware of an independent experiment arXiv:1904.12249.

**Acknowledgements:** This work was supported by the National Natural Science Foundation of China, the Chinese Academy of Sciences, and the National Fundamental Research Program.

**Figure captions**

**Fig. 1. Principle scheme for teleportation of three-dimensional quantum states**. Alice holds a quantum state $|\varphi\rangle_a$ encoded in three dimensions (depicted by three paths) that she wishes to teleport to Bob. To do so, they first share a three-dimensional, maximally entangled state. Then Alice performs a high-dimensional Bell-state measurement (HD-BSM) on her photons. Conceptually, our approach upon realising a HD-BSM consists of two parts: A unitary transformation in an expanded state space ($U_{3+1}$), and a multi-port beam splitter that enables collective quantum interference between Alice's teleportee photon ($a$), her part of the entangled state ($b$) and an additional ancillary photon ($x$). Specific "click-patterns" of different detectors indicate successful projections into one of the nine entangled Bell states. Alice can now transmit the classical-information of her "click-pattern" to Bob, who performs a unitary transformation ($U_3$) on his photon to recover the original state of Alice's teleportee photon.

**Fig. 2. Experimental set-up to teleport path-encoded qutrits**. An ultraviolet pulsed laser is used to create a three-dimensionally entangled photon-pair (path-encoded) in a non-linear

crystal (BBO$_1$) shared between Alice and Bob. The teleportee and ancillary photon are produced in a second non-linear crystal (BBO$_2$). All 12 input qutrit states to be teleported and the ancilla-photon are prepared using polarisation dependent beam-displacers (BDs) controlled by half- and quarter-wave plates (HWP, QWP). The expanded unitary transformation $U_{3+1}$ is implemented in a four-dimensional hybrid polarisation-path state space. A polarising beam splitter (PBS) traces out the additionally employed fourth-dimension. All three photons ($a$, $b$, $x$) enter the three-dimensional multi-port beamsplitter which consists of nested interferometers implemented in polarisation and path degree-of-freedom. A specifically designed partially polarising beam splitter (pPDBS) ensures equally distributed input ports to all output ports. Simultaneous "click-patterns" of detectors $\{a_0^{'}, a_1^{'}, a_2^{'}\}$, $\{b_0^{'}, b_1^{'}, b_2^{'}\}$ or $\{x_0^{'}, x_1^{'}, x_2^{'}\}$ indicates a successful BSM and heralds a teleported photon at Bob's side. No active feed-forward scheme was implemented here. Adjusting the HWP and QWP in Bob's measurement apparatus allows for a complete analysis of the teleported qutrits.

**Fig. 3. Experimental results of qutrit teleportation**. Measurement results for 6 out of all 12 basis states from different mutually unbiased bases groups $B_{\{1,2,3\}}^{(1-4)}$ for calculating the fidelities are displayed. Dashed empty bars indicate ideal measurement result for comparison. (**A**) All three computational basis states from the group $B_{\{1,2,3\}}^{(1)}$ and their relative four-photon occurrences are shown. (**B**)-(**D**) Measurement result of coherent superposition states from mutually unbiased bases groups $B_{\{1,2,3\}}^{(2)}$, $B_{\{1,2,3\}}^{(3)}$, and $B_{\{1,2,3\}}^{(4)}$ respectively. The different measurement outcomes $|\varphi_{ij}^{\pm}\rangle$ represent all possible two-dimensional combinations with phases according to the prepared qutrit state. Error bars are calculated using Monte-Carlo simulation with an underlying Poissonian count rate distribution.

**Fig. 4. Data summary for demonstrating a universal, non-classical and genuine qutrit teleportation experiment**. Here the fidelities of all 12 basis states are listed to demonstrate universality. Our average achieved fidelity of 0.75(1) significantly overcomes both, the non-classical bound of 1/2 and the genuine qutrit bound of 2/3. The shaded area is 1 standard deviation of the average fidelity. Error bars for fidelities are calculated with a Monte-Carlo simulation and an underlying Poissonian distribution of photon counts. Gaussian error propagation yields the error for the average fidelity.

**Figures:**

Fig. 1

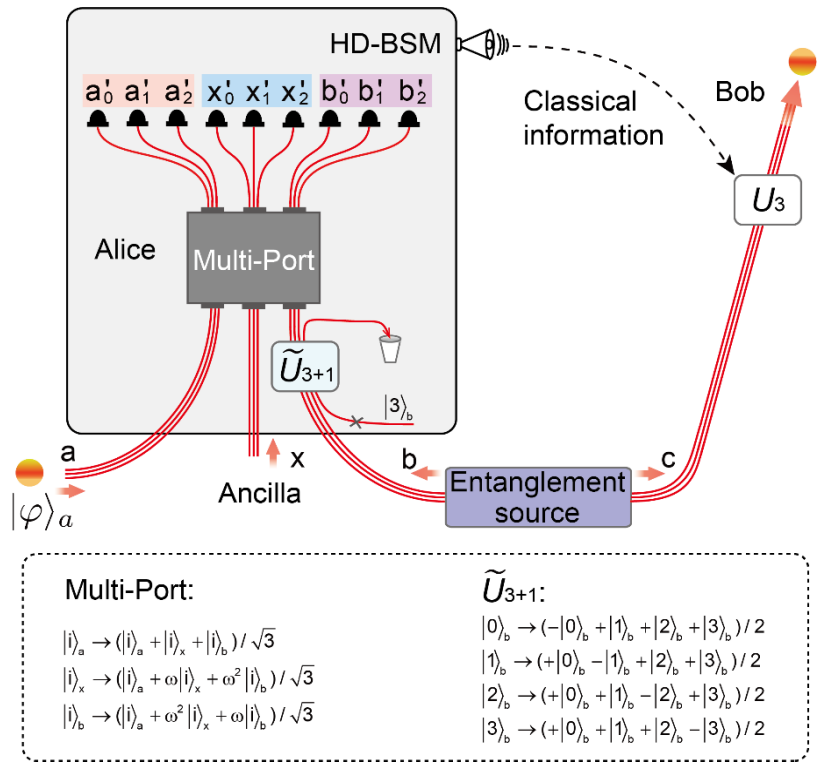

Fig. 2

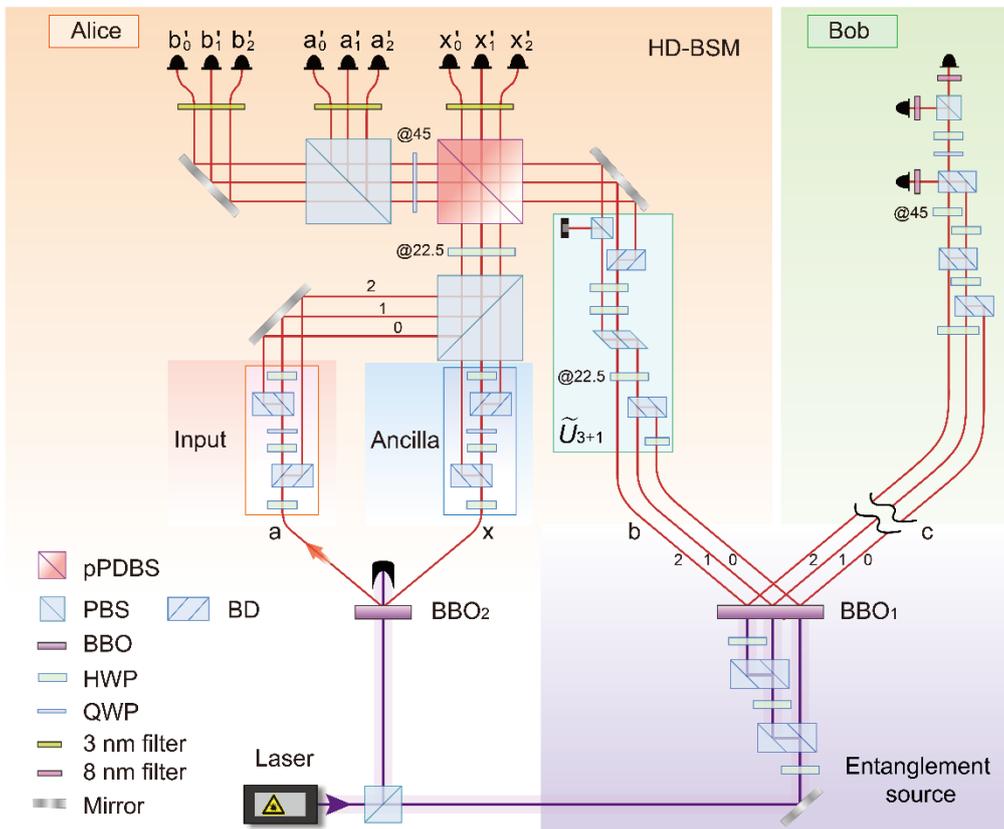

Fig. 3

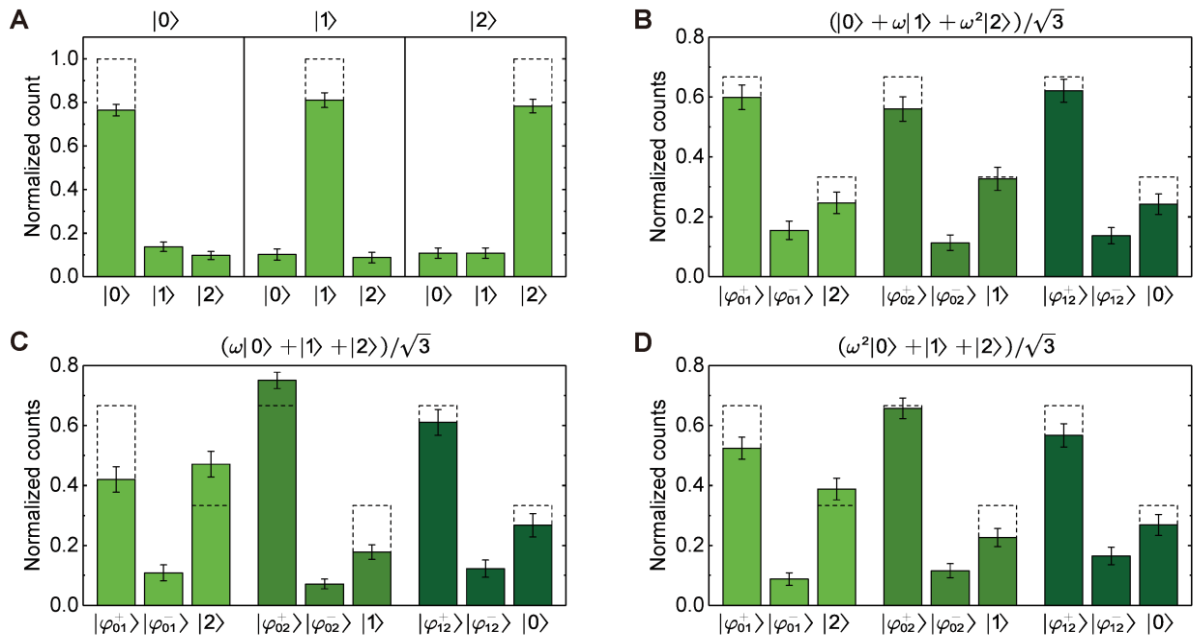

Fig. 4

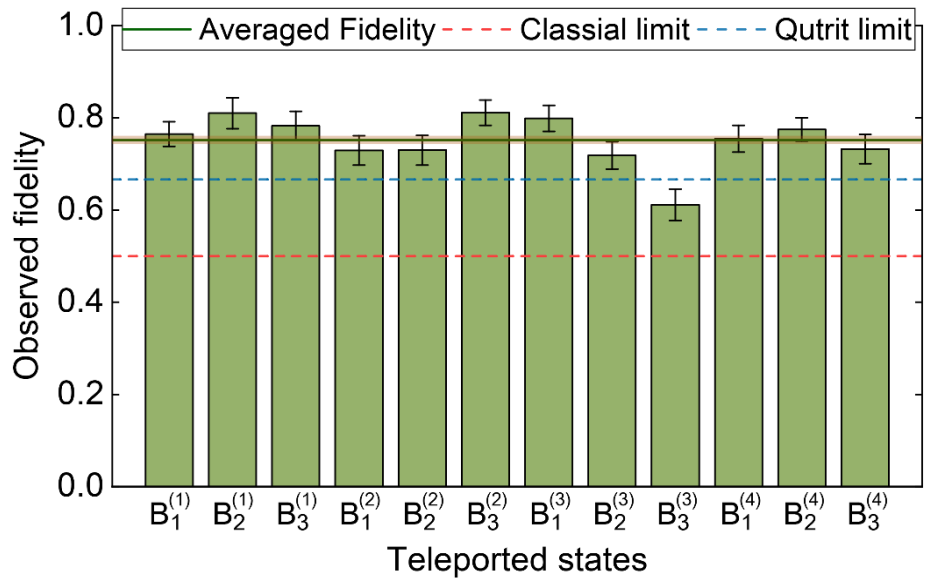

# Supplementary Information

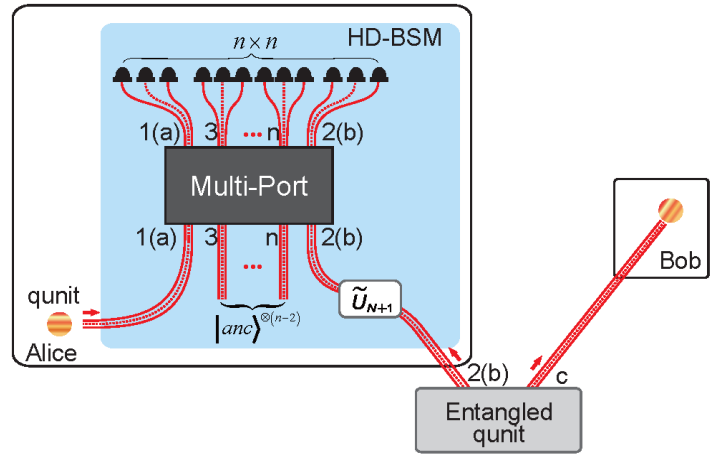

**Extended Data Figure 1 | Schematic of general N-dimensional quantum state teleportation using linear optics.** Alice and Bob initially share a maximally entangled quantum state in $N$-dimensions. The $N$-dimensional BSM is achieved by utilising $(N-2)$ additional ancilla photons $|\text{anc}\rangle$. An extended space unitary transformation $\tilde{U}_{N+1}$ in a $(N+1)$ dimensional state space with a $N$-dimensional quantum-fourier-transform that connects all input ports equally. A successful high-dimensional BSM is indicated by specific "click-patterns" that represent detection events of single photons in the $N \times N$ detectors. Alice anounces her "click-pattern" using a classical communication channel to Bob who applies a $N$-dimensional unitary transformation on his photon to retrieve the original $N$-dimensional teleportee photon.

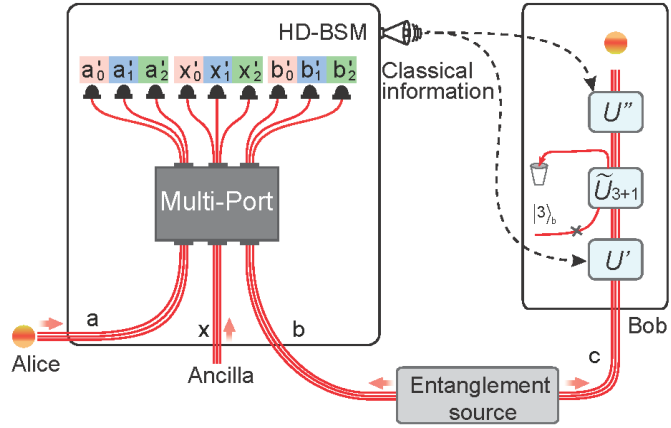

**Extended Data Figure 2 | Three-dimensional teleportation using active feed-forward.** Based on Figure 1 in main text, to enhance the success probability to $1/9$, the $\tilde{U}_{3+1}$ transformation is swapped to Bob's. In addition, two controlled unitary transformations $U'$ and $U''$ are also included. A successful teleportation is indicated by specific "click-patterns" that represent coincidence detection events of single photons in detectors marked with different subscript number $1, 2$ and $3$ (different colors). Alice transmits the classical-information of her "click-pattern to Bob, who selects the according parameters to perform proper operations $U'$ and $U''$, defined as equation (3) and (4). For each "click-pattern", the success probability of teleportation is $1/243$, and there are $3^3 = 27$ different "click-patterns" in total. Thus, the overall efficiency is $1/9$.

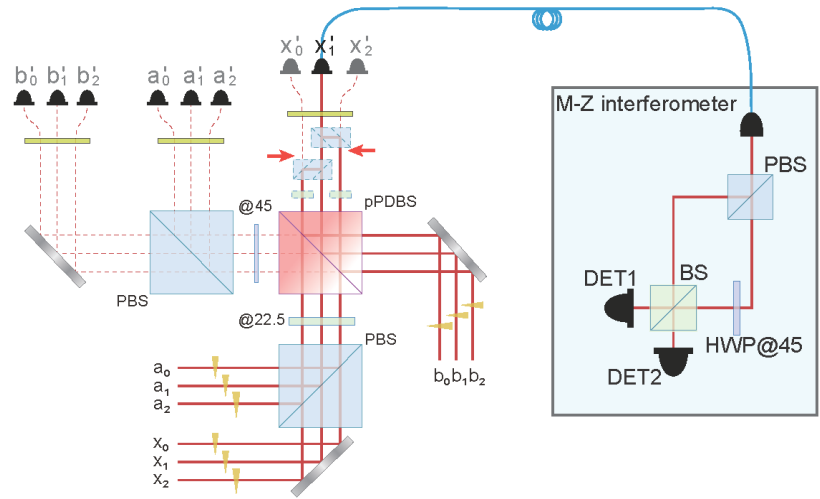

**Extended Data Figure 3 | Phase alignment details.** Experimental set-up to adjust all relative phases of all incoming paths for all input ports for the QFT. The pPDBS is used as a reference element in this alignment scheme. To align the relative phases between two paths (e.g. $x_1$ and $b_1$), their orthogonal polarisation is spatially overlapped using a BD and guided to a MZ interferometer. The MZ interferometer consists of a PBS, a HWP to flip the polarisation in one arm and a beam-splitter. The phase is referenced using the interference at the MZ interferometer by adjusting the optical wewedges (PS) in the respective paths.

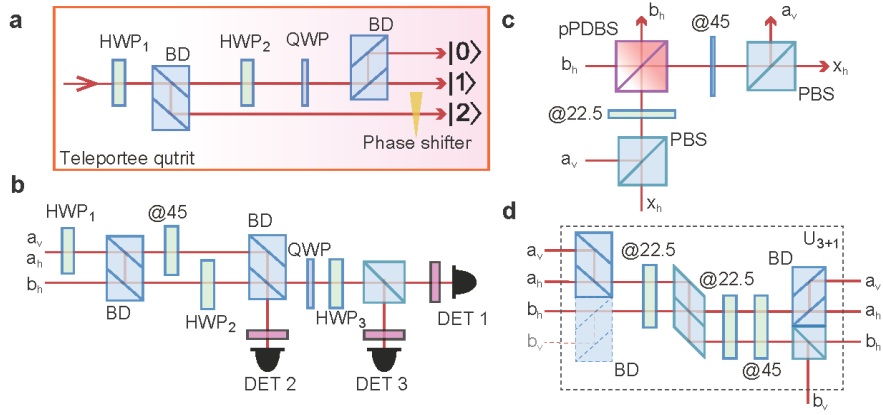

**Extended Data Figure 4 | Experimental details. a,** Detailed experimental set-up to prepare the teleportee photon. Here, the polarisation of the teleportee photon is utilised. HWP, QWP and PS in combination with polarisation sensitive beam-displacers are utilised to create all 12 states to be teleported. **b,** Detailed experimental realisation of Bob's measurement apparatus: In analogy to the teleportee photon prepartation, the polarisation is used to first transfer Bob's photon into the eigenstates of our measurement operators. Three single-photon detectors (DET1, DET2, DET3) allow for multi-outcome measurements. **c,** Physical realisation of the three-dimensional QFT in a hybrid polarisation-path state space using a specifically designed pPDBS, two PBS in combination with a HWP at $22.5°$ and a QWP at $45°$. The polarisation state of all input and output ports are denoted as $(h, v)$ for horizontal and vertical, respectively. **d,** Detailed depiction of the extended space unitary $\bar{U}_{3+1}$. The added dimension $b_v$ is illustrated with dotted lines to clarify that this input port is empty, meaning no photon-path is connected here.

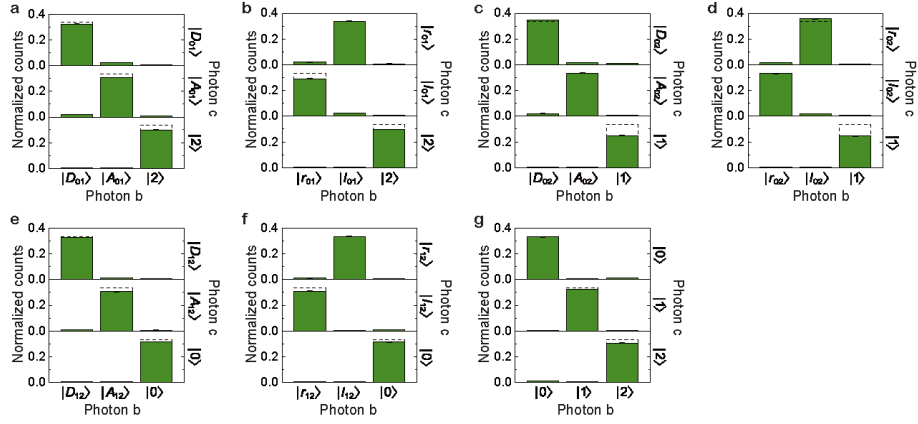

**Extended Data Figure 5 | Detailed measurements for the three-dimensionally entangled source.** In our experiment, we create a three-dimensionally entangled photon-pair $b$ and $c$. **a-g** correspond to measurement result of observable $\sigma^x_{01} \otimes \sigma^x_{01}$, $\sigma^y_{01} \otimes \sigma^y_{01}$, $\sigma^x_{02} \otimes \sigma^x_{02}$, $\sigma^y_{02} \otimes \sigma^y_{02}$, $\sigma^x_{12} \otimes \sigma^x_{12}$, $\sigma^y_{12} \otimes \sigma^y_{12}$ and population $\mathcal{P}$ respectively. Fidelity is then calculated with expectations of the seven observables. For each observable, there are three possible result for both photon $b$ and photon $c$, here we list the 9 normalized counts (correspond to nine combination of photon $b$ and $c$) in $3 \times 3$ shape. In this figure, $|D_{ij}\rangle = (|i\rangle + |j\rangle)/\sqrt{2}$, $|A_{ij}\rangle = (|i\rangle - |j\rangle)/\sqrt{2}$, $|r_{ij}\rangle = (|i\rangle + i|j\rangle)/\sqrt{2}$, $|l_{ij}\rangle = (|i\rangle - i|j\rangle)/\sqrt{2}$.

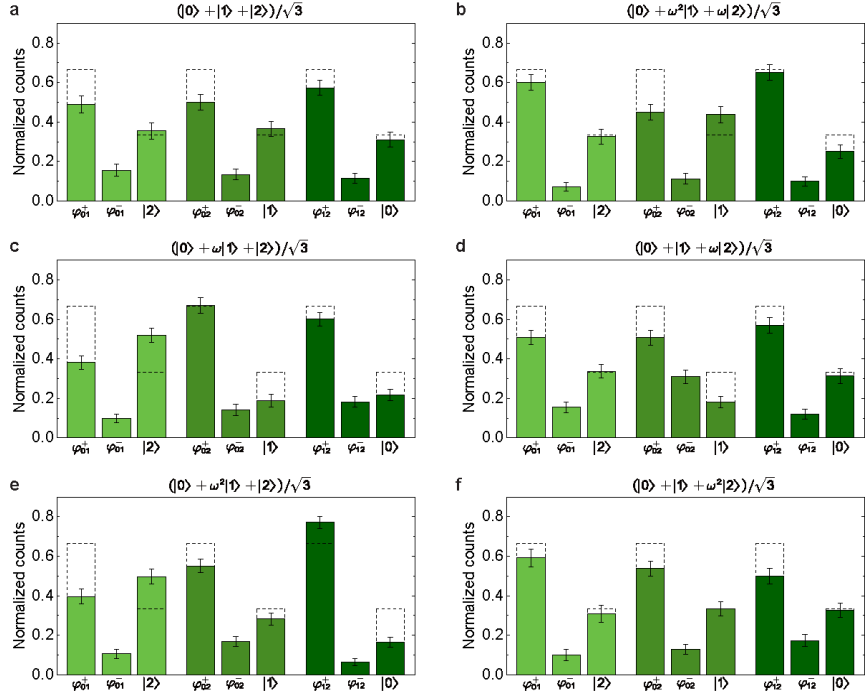

**Extended Data Figure 6 | Extended data of teleported states.** Measurement results of the remaining 6 basis states (out of 12) from different MUBs sets ($B^{(1)}$, $B^{(2)}$, $B^{(3)}$, $B^{(4)}$) for calculating the fidelities are displayed. Dashed empty bars indicate ideal measurement result for comparison. The different measurement outcomes $|\varphi_{ij}^{\pm}\rangle$ represent all possible two-dimensional combinations with phases according to the prepared qutrit state. Error bars are calculated using Monte-Carlo simulation with an underlying Poissonian count rate distribution.

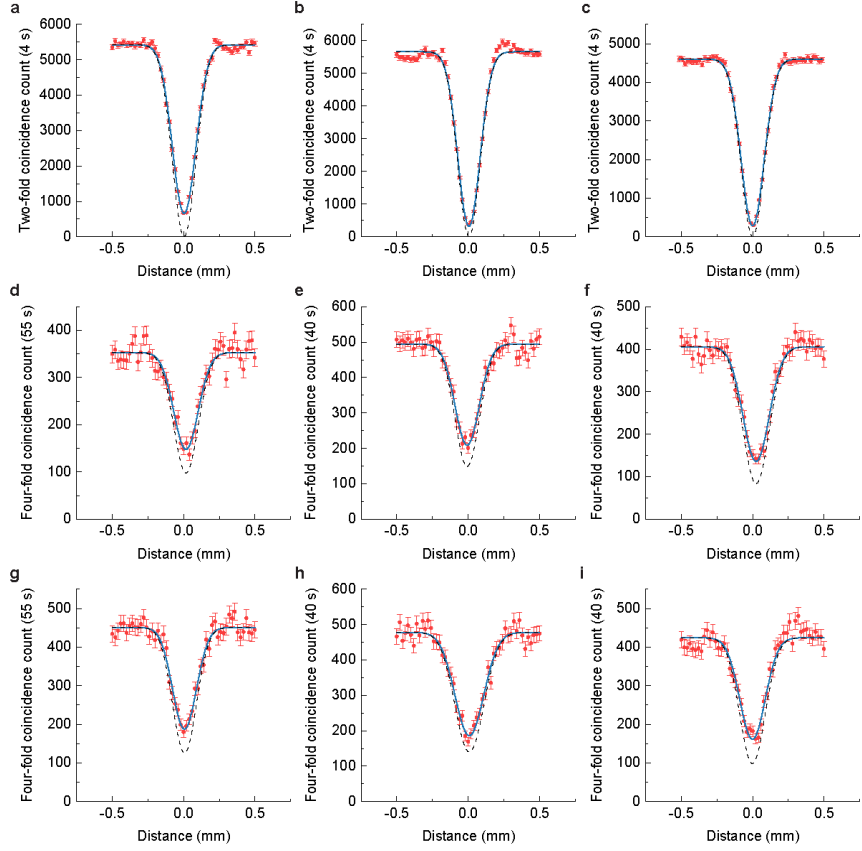

**Extended Data Figure 7 | Hong-Ou-Mandel interference for all input ports and paths at the QFT. a-c,** Interference curve of mode $0$, $1$ and $2$ between ancillary photon and teleportee photon. The two photons are prepared to $45°$ polarization, then combined on PBS. The curve is obtained by scanning "trombone" system 1 for ancilla photon. For the two photons are generated from the same BBO crystal, influence of high-order double pair emission is minor, consistent with the high visibility observed ($92\%$ on average). **d-f,** HOM interference curve of mode $0$, $1$ and $2$ between teleportee photon and photon b; **g-i,** HOM interference curve of mode $0$, $1$ and $2$ between ancilla photon and photon b. The data displayed in **d-i** demonstrates the interplay between photons generated from two separate BBO crystals with narrow-band filters (3 nm). The typical HOM curves are obtained by scanning "trombone" system 1 and 2 for the ancilla and teleportee photon respectively. The dashed line shows the theoretically expected ideal situation. To guide the eye, we also fit a gaussian function to the measured data points. On average, the visibility is $82\%$. **g-i,** The observed average width and visibility coincides with theoretical expectation due to the chosen narrow band filters spectral filters ($3+8$ nm for **a-c**, $3+3$ nm for **d-i**). Error bars are due to poissonian counting statistics.

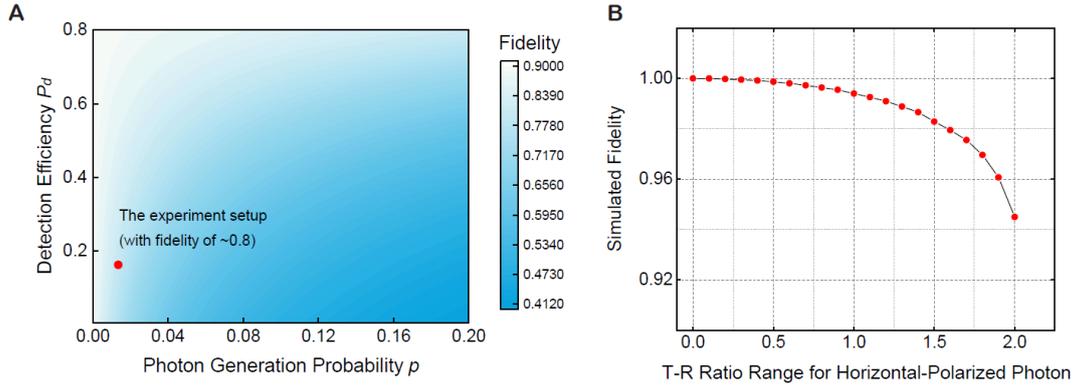

**Fig. S8. Numerical analysis of the teleportation fidelity. A.** Calculated fidelitiy for teleportation with different detection efficiency $P_d$ (including efficiency for photon collection and detector) and photon generation probability $p$. For the system used in our experiment, $P_d \approx 0.16$, $p \approx 0.013$, the estimated fidelity $F \sim 0.8$ can be read from the figure, which is consistent with the estimation according to measured HOM interference visibility. **B.** Influence of imperfect splitting ratio. In experiment, the used pPDBSs should ideally have transmission-reflection ratio of $2 : 1$ for horizontal-polarized photon. Experimentally, there is always a deviation from the ideal value. We simulate the influence by adding the ratio $2 : 1$ a random variation for each input modes. Each point shown in the figure is the mean value of 1000 times caculation with a specific range of variation corresponding to $x$-axis value. Our numerical analysis shows that the devitation of the reflectivity only slightly influence the teleportation fidelity.